\documentclass{PoS}

\usepackage{subfig}
\usepackage{cite}
\usepackage{graphicx}
\usepackage{amsmath}
\title{Antenna Subtraction in pQCD at NNLO}

\ShortTitle{Antenna Subtraction in pQCD at NNLO}

\author{\speaker{James Currie}\\
        Institute for Particle Physics Phenomenology\\
        Department of Physics\\
        University of Durham\\
        DH1 3LE\\
        UK\\
        E-mail: \email{j.r.currie@durham.ac.uk}\\
        }

\abstract{In this talk I discuss the antenna subtraction method for isolating infrared (IR) singularities of jet cross sections in perturbative QCD.  The method is applied at next-to-next-to-leading order (NNLO) to dijet production in hadron collisions at the LHC.  The double real radiative corrections to the dijet cross section are considered and their IR behaviour is examined.  IR subtraction terms are constructed to absorb numerical divergences of the cross section in the single and double unresolved regions of phase space using the antenna subtraction method. A pictorial representation of matrix elements and antenna functions is presented with specific examples of how such diagrams can be used in practical calculations.}

\FullConference{10th International Symposium on Radiative Corrections (Applications of Quantum Field Theory to Phenomenology)\\
		 September 26-30, 2011\\
		 Mamallapuram, India}

\begin{document}

\section{Introduction}

One of the most powerful tools for describing the physics of high energy hadronic collisions is perturbative QCD in the context of the parton model.  In this framework, the cross section for two partons to scatter into a number of jets can be calculated as a perturbative expansion in the strong coupling, $\alpha_{s}$~\cite{pinkbook},
\begin{eqnarray}
\textnormal{d}\sigma&=&\sum_{i,j}\int\ \frac{\textnormal{d}\xi_{1}}{\xi_{1}}\frac{\textnormal{d}\xi_{2}}{\xi_{2}}\ f_{i}(\xi_{1})f_{j}(\xi_{2})\ \biggl[\ \textnormal{d}\hat{\sigma}_{ij}^{LO}+\bigg(\frac{\alpha_{s}}{2\pi}\biggr)\textnormal{d}\hat{\sigma}_{ij}^{NLO}+\bigg(\frac{\alpha_{s}}{2\pi}\biggr)^{2}\textnormal{d}\hat{\sigma}_{ij}^{NNLO}+{\cal{O}}(\alpha_{s}^{3})\ \biggr]
\end{eqnarray}
where $\xi_{1},\xi_{2}$ and $f_{i}(\xi_{1}),f_{j}(\xi_{2})$ denote the momentum fraction and Parton Density Function (PDF) associated with initial state parton $i$ and $j$ respectively. By calculating the higher order contributions, the theoretical prediction for the cross section can be improved in several respects~\cite{nigel2002}.  In high energy collisions where $\alpha_{s}$ is small, higher order corrections naturally improve the convergence of the perturbative series and significantly reduce the renormalization and factorization scale uncertainty.  The partonic cross section can only be compared with experimentally observed jets by use of a jet clustering algorithm.  Such algorithms benefit from having more particles in the final state and so higher order calculations allow a more realistic comparison between parton level and hadron level jets~\cite{salam}.  Similarly higher order corrections allow for multiple initial state radiation which in turn generates transverse momentum for the final state.
\par Next-to-leading order (NLO) calculations of the dijet cross section have been available for some time and have been tested against experimental data at both the TEVATRON and the LHC~\cite{NLOdijet,dijetdiff}. Currently the cutting edge for jet cross sections is the NNLO contribution to dijet production with hadronic initial states.  NNLO calculations contain several components: the two loop double-virtual cross section, $\textnormal{d}\hat{\sigma}_{NNLO}^{VV}$, the one loop real-virtual contribution, $\textnormal{d}\hat{\sigma}_{NNLO}^{RV}$ and the double-real tree level contribution, $\textnormal{d}\hat{\sigma}_{NNLO}^{RR}$.  Each component contributes to the same order of $\alpha_{s}$ and so the terms with a lower number of loops have a higher number of final state particles thus requiring integration over a larger dimensional phase space.  The dijet cross section can be written in terms of these quantities and expressed as~\cite{gluons},
\begin{eqnarray}
\textnormal{d}\sigma_{NNLO}&=&\int_{\Phi_{4}}\ \textnormal{d}\hat{\sigma}_{NNLO}^{RR}+\int_{\Phi_{3}}\ \textnormal{d}\hat{\sigma}_{NNLO}^{RV}+\int_{\Phi_{2}}\ \textnormal{d}\hat{\sigma}_{NNLO}^{VV}
\label{xsec}
\end{eqnarray}
where the requirement that exactly two jets are observed is implemented by the jet function, $J^{(n)}_{2}$ which is defined to construct two jets from $n$ final state partons,
\begin{eqnarray}
\int_{\Phi_{n}}&=&\int\ \textnormal{d}\Phi_{n}\ J^{(n)}_{2}(p_{1},\cdots,p_{n}).
\end{eqnarray}
\indent The calculation of each of the three terms in Eq. \eqref{xsec}  presents different challenges:  The double-virtual contribution has only explicit poles arising directly from the loop integrations but the loop integration itself is highly non-trivial. On the other hand the matrix elements for the double-real contribution have been known for many years but their intricate implicit IR singularity structure and more complicated phase space integration has caused a bottleneck in the calculation.  The real-virtual contribution contains both implicit and explicit IR singularities.  In this talk we will focus on the double-real radiative corrections to the dijet cross section for processes involving quarks and gluons in either the initial or final state. IR divergences are inherent to the double-real cross section and arise from field configurations where partons become soft or collinear in any combination.  The factorization of the colour ordered matrix elements in these IR configurations is universal~\cite{factorize} and this is exploited to isolate the singular behaviour of the cross section.\vspace{-0.5cm}

\section{Antenna Subtraction}

A significant effort has been applied to the calculation of the NNLO dijet cross section~\cite{Kosower,gluongluon,quarkgluon,initialnlo,luisoni,ritzmann,monni} and all necessary matrix elements are now known.  The remaining challenge is to isolate the implicit (local) IR singularities in the double-real and real-virtual contributions, to explicitly cancel the IR poles between terms and to numerically integrate the resultant finite integrand over the appropriate phase space.  Many techniques have been developed for the purpose of isolating IR poles both at NLO and NNLO~\cite{dipole,FKS,sector,NNLOantenna,stripper}.  The antenna subtraction method~\cite{NNLOantenna} is one such formalism and has been applied successfully at NNLO to the calculation of $e^{+}e^{-}\rightarrow3j$~\cite{ee3j,Weinzierl}.  The method consists of constructing a subtraction term, $\textnormal{d}\hat{\sigma}_{NNLO}^{S}$, from antenna functions and reduced multiplicity matrix elements which mimics the IR behaviour of the physical matrix elements in all singular configurations.  A subtraction term is also constructed to absorb the implicit IR singularities of the one loop real-virtual contribution.  With the subtraction terms defined properly and including mass factorization terms, $\textnormal{d}\sigma_{NNLO}^{MF,1,2}$, the cross section may be rewritten in the form~\cite{gluons},
\begin{eqnarray}
\textnormal{d}\sigma_{NNLO}&=&\int_{\Phi_{4}}\ \biggl[\textnormal{d}\hat{\sigma}_{NNLO}^{RR}-\textnormal{d}\hat{\sigma}_{NNLO}^{S}\biggr]+\int_{\Phi_{4}}\ \textnormal{d}\hat{\sigma}_{NNLO}^{S}\nonumber\\
&+&\int_{\Phi_{3}}\ \biggl[\textnormal{d}\hat{\sigma}_{NNLO}^{RV}-\textnormal{d}\hat{\sigma}_{NNLO}^{VS}\biggr]+\int_{\Phi_{3}}\ \textnormal{d}\hat{\sigma}_{NNLO}^{VS}+\textnormal{d}\hat{\sigma}_{NNLO}^{MF,1}\nonumber\\
&+&\int_{\Phi_{2}}\ \textnormal{d}\hat{\sigma}_{NNLO}^{VV}+\textnormal{d}\hat{\sigma}_{NNLO}^{MF,2}.
\end{eqnarray}
In doing so, the difference between the double-real emission matrix elements and the subtraction term is rendered IR finite and the integrand can be integrated over the four-parton phase space, yielding a finite result.  The subtraction term is constructed in such a way  to allow for analytic integration of the antenna function containing all IR divergences over the antenna phase space, a subspace depending only on the parton momenta involved in a singular configuration.  Upon integration in $d=4-2\epsilon$ dimensions the IR poles may be extracted, cancelled against the explicit poles coming from the virtual amplitudes and the finite remainder numerically integrated.  The final goal is to rewrite the cross section in the form,
\begin{eqnarray}
\textnormal{d}\sigma_{NNLO}&=&\int_{\Phi_{4}}\ \biggl[\textnormal{d}\hat{\sigma}_{NNLO}^{RR}-\textnormal{d}\hat{\sigma}_{NNLO}^{S}\biggr]\nonumber\\
&+&\int_{\Phi_{3}}\ \biggl[\textnormal{d}\hat{\sigma}_{NNLO}^{RV}-\textnormal{d}\hat{\sigma}_{NNLO}^{T}\biggr]\nonumber\\
&+&\int_{\Phi_{2}}\ \biggl[\textnormal{d}\hat{\sigma}_{NNLO}^{VV}-\textnormal{d}\hat{\sigma}_{NNLO}^{U}\biggr],
\end{eqnarray}
where each term is individually finite and can be numerically integrated over the appropriate phase space in four dimensions,
\begin{eqnarray}
\textnormal{d}\hat{\sigma}_{NNLO}^{T}&=&\textnormal{d}\hat{\sigma}_{NNLO}^{VS}-\textnormal{d}\hat{\sigma}_{NNLO}^{MF,1}-\int_{\chi_{1}}\ \textnormal{d}\hat{\sigma}_{NNLO}^{S}\nonumber\\
\textnormal{d}\hat{\sigma}_{NNLO}^{U}&=&-\textnormal{d}\hat{\sigma}_{NNLO}^{MF,2}-\int_{\chi_{1}}\ \textnormal{d}\hat{\sigma}_{NNLO}^{VS}-\int_{\chi_{2}}\ \textnormal{d}\hat{\sigma}_{NNLO}^{S}
\end{eqnarray}
where $\chi_{n}$ denotes the phase space corresponding to the parton momenta involved in a singular configuration with $n$ unresolved partons.

\par In the antenna formalism, the subtraction term is constructed from an antenna function containing all the IR singular behaviour and a finite reduced matrix element.  The antenna function's arguments are the momenta of two hard radiators and one or two momenta that are allowed to become unresolved.  The arguments of the matrix element used for the subtraction term are obtained from the original momenta via a phase space map~\cite{Kosowermap,initialnlo}.  The momenta involved in the antenna function are mapped onto two composite momenta such that the matrix element is independent of the unresolved momenta in the singular limits.  The subtraction term for a double real emission $n$ parton process can be schematically represented as a sum over terms of the form,
\begin{eqnarray}
\textnormal{d}\hat{\sigma}_{NNLO}^{S}&\sim&X_{m+2}(i_{1},\cdots,i_{m+2})\ \otimes\ M_{n-m}^{0}(j_{1},\cdots,j_{n-m})
\end{eqnarray}
where $X_{m+2}$ is a $(m+2)$ parton antenna function containing at most $m$ unresolved partons and $M_{n-m}^{0}$ is a squared matrix element summed over colours and helicities.  At NNLO we are interested in three and four parton antenna functions to absorb the single and double unresolved divergences.  We consider colour ordered amplitudes because of their universal factorization properties in unresolved limits; the subtraction term is constructed to capture this factorized behaviour.
\begin{figure}[h]
\centering
\subfloat[$X_{6}$ ]{\includegraphics[width=1.7cm]{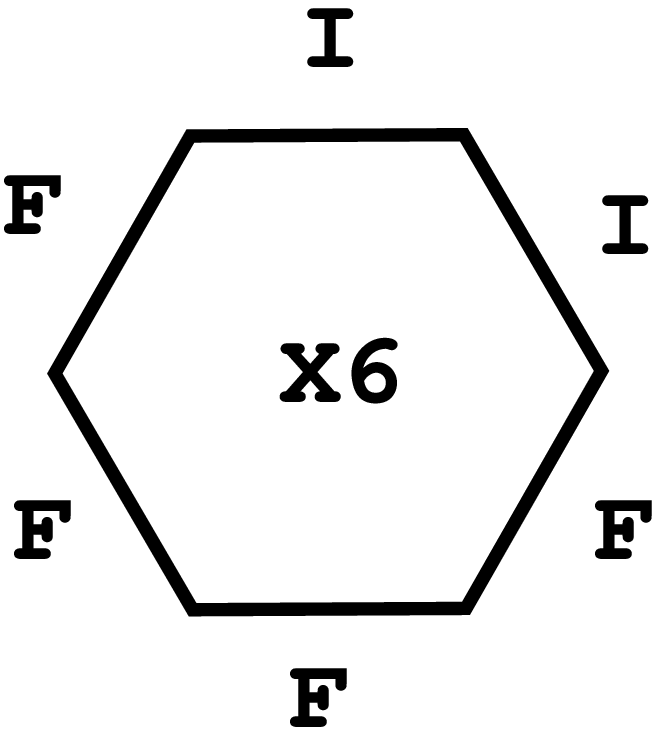}}\hspace{2cm}
\subfloat[$Y_{6}$ ]{\includegraphics[width=1.7cm]{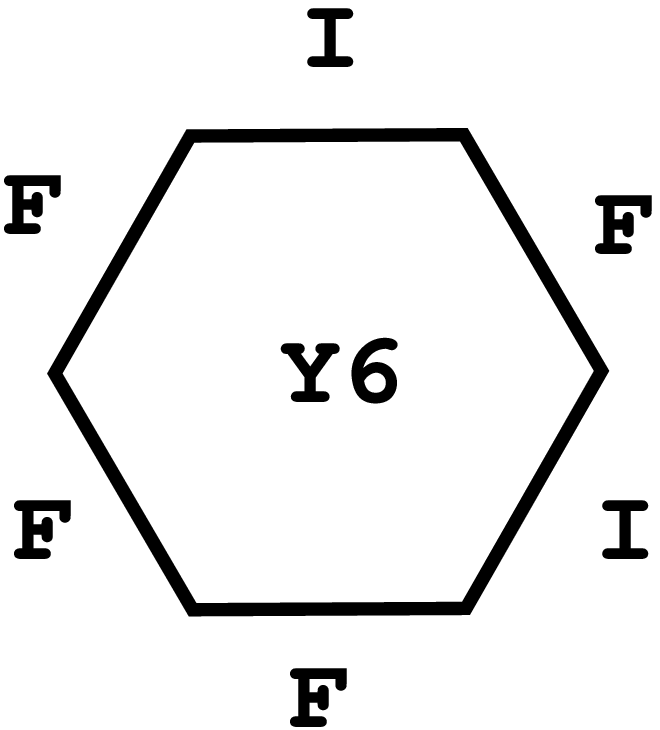}}\hspace{2cm}
\subfloat[$Z_{6}$ ]{\includegraphics[width=1.7cm]{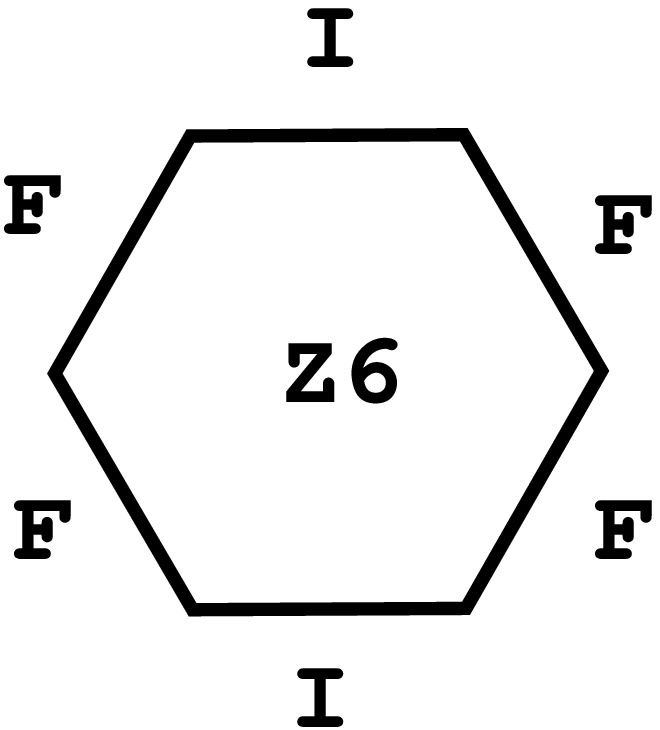}}
\caption{\footnotesize{The three hexagon topologies for $2\rightarrow4$ scattering. \emph{I} and \emph{F} denote initial and final state edges respectively.}}
\label{hextop}
\end{figure}
\par A novel way to understand how to construct subtraction terms is to consider the problem diagrammatically.  Consider a $n$-parton squared matrix element as a $n$-gon with each edge representing the momenta of an external leg.  For six parton scattering the real matrix elements are represented as hexagons as in Fig. \ref{hextop}.  Two of the six edges are labelled with an \emph{I} to denote that these edges represent initial state partons.  There are three independent ways two \emph{I}'s can be distributed about the hexagon, defining three kinematic topologies, $X_{6},Y_{6},Z_{6}$, see Fig. \ref{hextop}.
\par
\begin{figure}[h]
\centering
\subfloat[$X_{4}$ ]{\includegraphics[width=1.5cm]{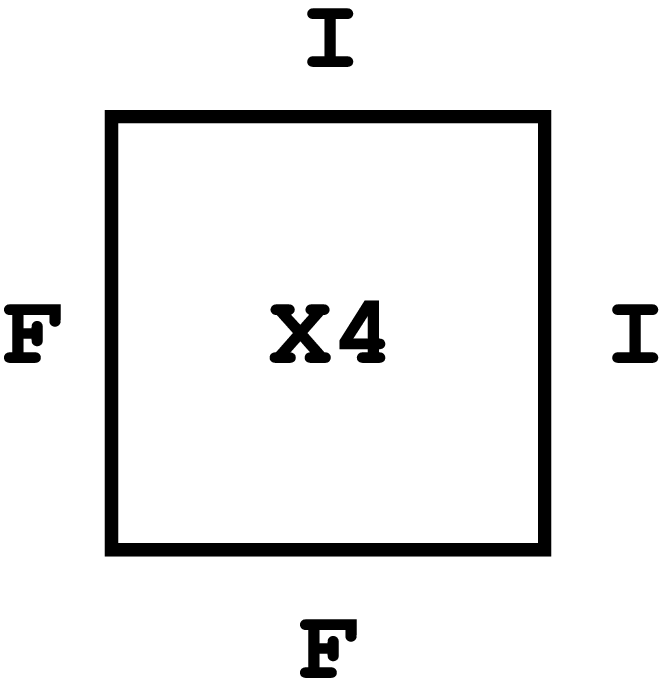}}\hspace{1cm}
\subfloat[$Y_{4}$ ]{\includegraphics[width=1.5cm]{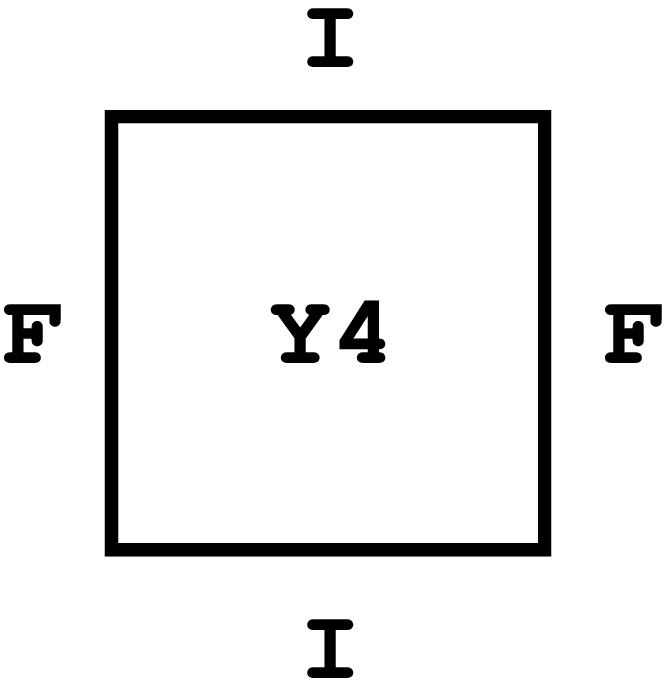}}\hspace{1cm}
\subfloat[$X_{5}$ ]{\includegraphics[width=1.5cm]{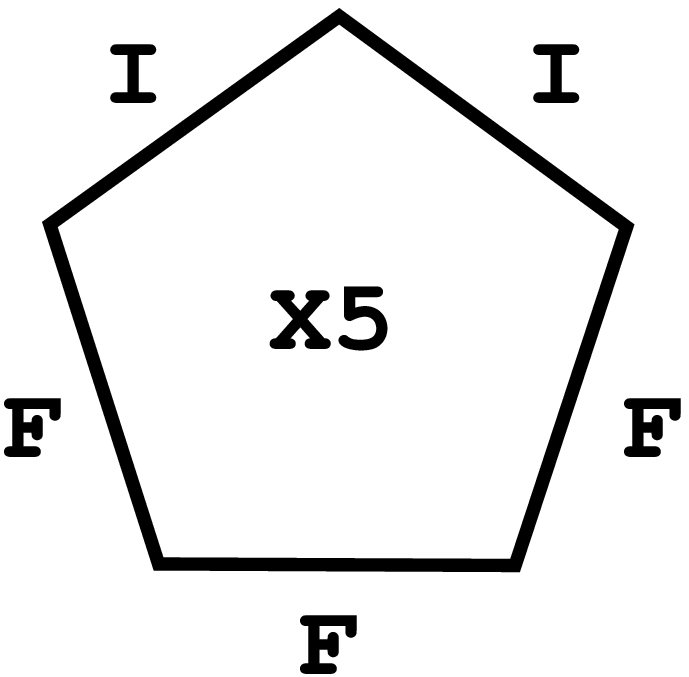}}\hspace{1cm}
\subfloat[$Y_{5}$ ]{\includegraphics[width=1.5cm]{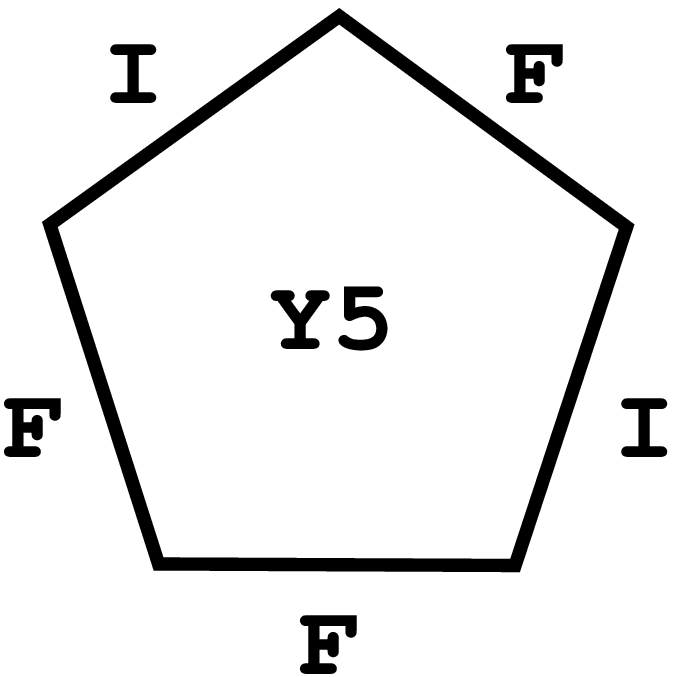}}
\caption{\footnotesize{The two box and pentagon topologies representing four and five parton matrix elements.}}
\label{boxpent}
\end{figure}

 As well as the six parton matrix elements, the calculation involves four and five parton matrix elements in the subtraction terms.  These are represented as box and pentagon diagrams and displayed in Fig. \ref{boxpent}.  In these cases there are only two independent kinematic topologies, $X_{4},Y_{4}$ and $X_{5},Y_{5}$ for the boxes and pentagons respectively.
\par
\begin{figure}[h]
\centering
\subfloat[$X_{3}$ ]{\includegraphics[width=1.3cm]{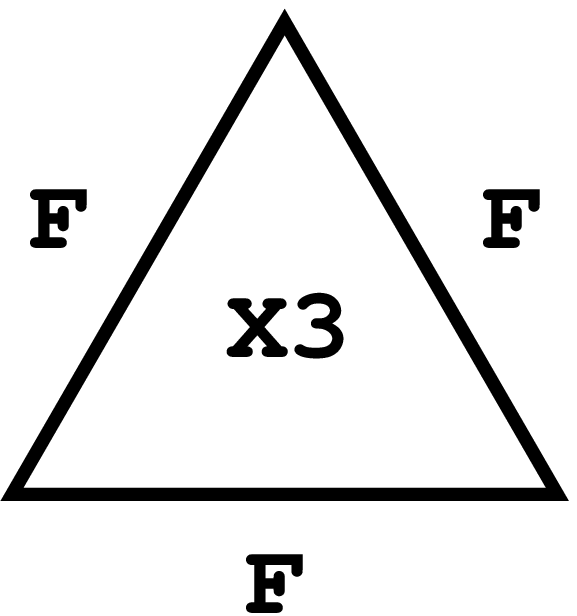}}\hspace{1cm}
\subfloat[$Y_{3}$ ]{\includegraphics[width=1.3cm]{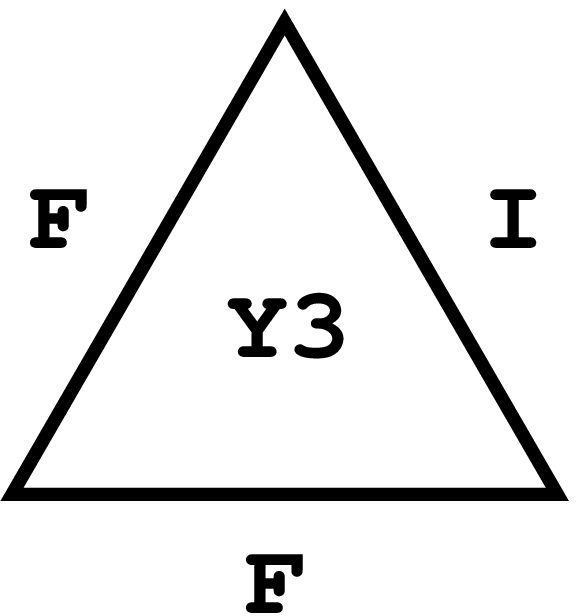}}\hspace{1cm}
\subfloat[$Z_{3}$ ]{\includegraphics[width=1.3cm]{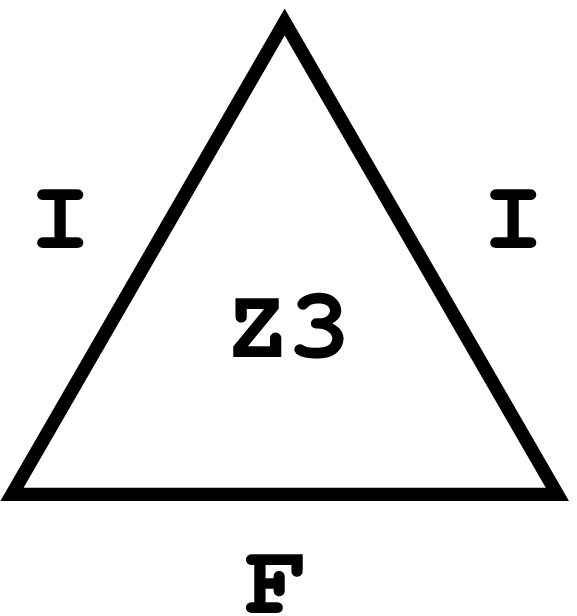}}\hspace{1cm}
\subfloat[$V_{4}$ ]{\includegraphics[width=1.5cm]{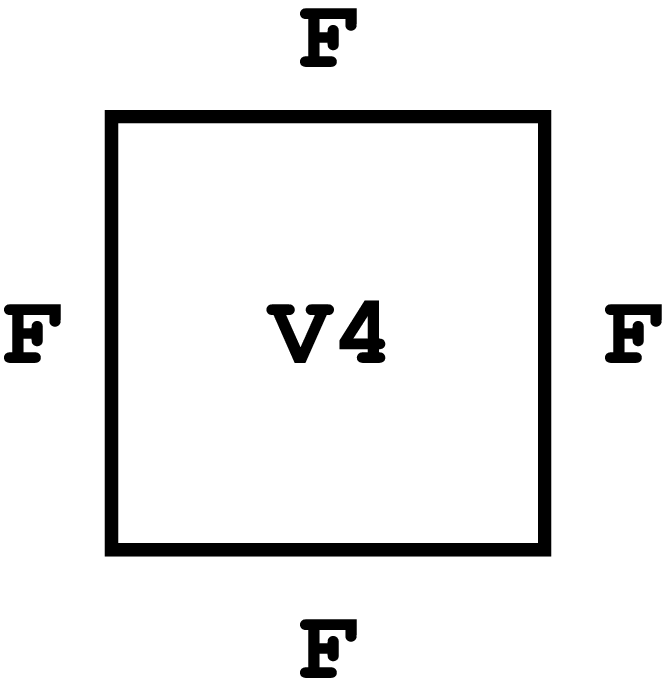}}\hspace{1cm}
\subfloat[$W_{4}$ ]{\includegraphics[width=1.5cm]{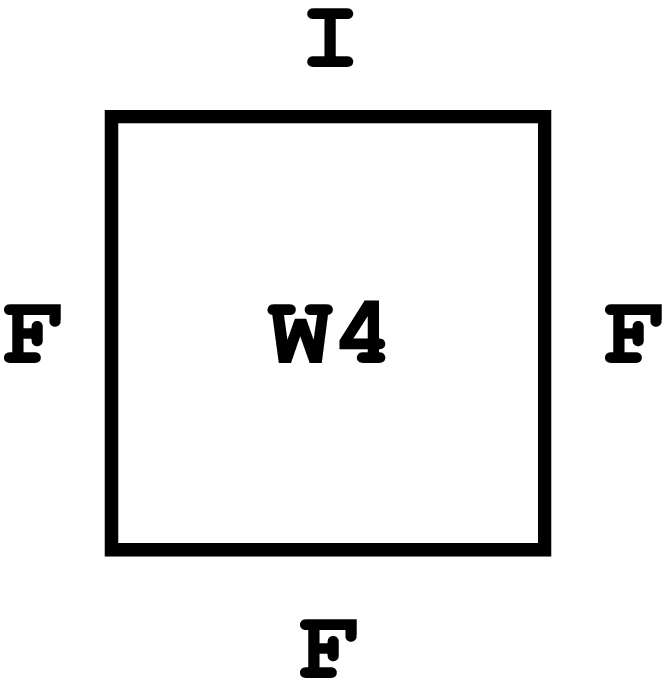}}
\caption{\footnotesize{The three triangle and two additional box topologies used to represent the three and four parton antenna functions.}}
\label{tribox}
\end{figure}

 The antenna functions which reflect the IR behaviour of squared matrix elements and are colour ordered can also be represented in pictorially.  Three and four parton antennae are represented as triangles and boxes where now the number of initial state edges can be zero, one or two, depending on whether we are representing a final-final, initial-final or initial-initial antenna, see Fig. \ref{tribox}.
\par
\begin{figure}[h]
\centering
\includegraphics[width=11cm]{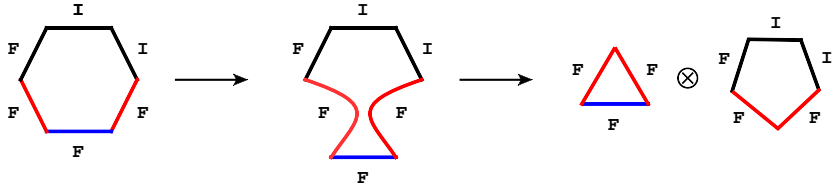}
\caption{\footnotesize{A diagrammatic representation of single unresolved parton factorization for six parton scattering belonging to the $X_{6}$ topology.  The unresolved edge (blue) is adjacent to two hard edges (red).  Edges are labelled \emph{I} and \emph{F} to denote initial or final state partons.}}
\label{pinch}
\end{figure}

 Using this diagrammatic approach, the factorization of the matrix elements is quite intuitive and follows a diagrammatic algorithm, represented in Fig. \ref{pinch}.  First consider a final state parton becoming unresolved and identify its hard colour adjacent neighbours.  The hard neighbouring edges are then stretched and pinched to a point, separating the original polygon into two daughter polygons.  The daughter polygon containing the unresolved edges represents the appropriate antenna function and the remaining polygon containing only resolved edges represents the reduced matrix element with mapped momenta.  

\par
\begin{figure}[h]
\centering
\includegraphics[width=9.5cm]{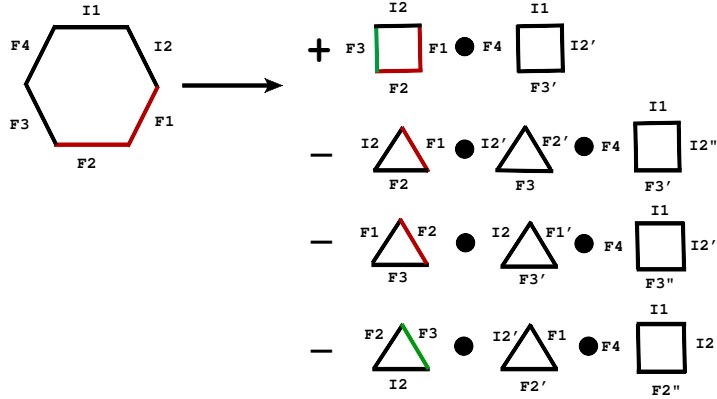}
\caption{\footnotesize{The block of subtraction terms used to absorb divergences arising from colour connected partons $F_{1}$ and $F_{2}$ (blue) becoming unresolved in a six parton matrix element (hexagon) belonging to the $X_{6}$ topology.  Colour connected double unresolved limits are removed using the four parton antenna (box). Spurious single unresolved limits of the box are removed using iterated three parton antennae (triangles) including the extra limit when $F_{3}$ becomes unresolved (green).}}
\label{w4block}
\end{figure}

 The unresolved limits of the six parton matrix elements can be systematically examined by simply taking all the possible pinches of the appropriate hexagon.  In the single unresolved limits the hexagon is pinched to give a product of a triangle and a pentagon, see Fig. \ref{pinch}.  A colour connected double unresolved limit is obtained by letting two adjacent edges go unresolved; in this case the hexagon is pinched to form the product of two boxes.  Almost colour connected limits are derived by iteratively pinching out two edges separated by a single hard edge, yielding a product of two triangles with one edge in common and a box.  Colour disconnected double unresolved limits are taken by pinching out two edges with no neighbours in common resulting in the product two triangles with distinct edges and a box.  The four parton antenna functions also contain single unresolved limits of their own which have to be removed.  This is achieved by applying the same algorithm to the box as was done for the hexagon; single unresolved edges are pinched out and subtraction terms are constructed from a product of two triangles.  In this way the full subtraction term can be constructed from blocks of subtraction terms which remove unresolved limits of the matrix elements in a systematic way whilst not introducing spurious singularities of their own.  In Fig. \ref{w4block} a block is defined to remove a specific double unresolved limit from the six parton matrix element using a four parton antenna function whose spurious single unresolved limits are removed in turn by products of triangles.
\par
\begin{figure}[h]
\centering
\subfloat[]{\includegraphics[width=1.8cm]{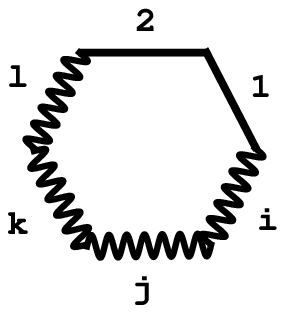}}\hspace{2cm}
\subfloat[]{\includegraphics[width=1.5cm]{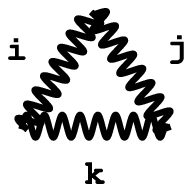}}\hspace{2cm}
\subfloat[]{\includegraphics[width=2.1cm]{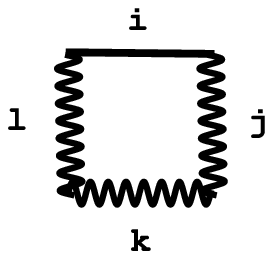}}
\caption{\footnotesize{Diagrams can also be drawn with wavy edges to represent gluons and straight edges to represent quarks for either matrix elements or antenna functions. (a) $M_{6}^{0}(\hat{1}_{q},i,j,k,l,\hat{2}_{\bar{q}})$, (b) $F_{3}^{0}(i,j,k)$, (c) $D_{4}^{0}(i_{q},j,k,l)$ }}
\label{newedges}
\end{figure}

 The specific antenna functions and matrix elements used in this block depend on the partonic channel under examination and can be read off directly from the diagrams when the edges are labelled with the appropriate parton species.  In six gluon scattering, the block of subtraction terms in Fig. \ref{w4block} absorbs divergences coming from the $i,j$ unresolved limit of the matrix element $M_{6}^{0}(\hat{1},\hat{2},i,j,k,l)$, where the hat denotes initial state partons\footnote{This set of matrix elements is represented diagrammatically by a hexagon belonging to the $X_{6}$ topology}. Translating the block of diagrams from Fig. \ref{w4block} into the specific functions for this process yields the subtraction term~\cite{gluons},
\begin{eqnarray}
F_{4}^{0}(\hat{2},i,j,k)\ M_{4}^{0}(\hat{1},\hat{\bar{2}},\widetilde{(ijk)},l)&-& f_{3}^{0}(\hat{2},i,j)\ F_{3}^{0}(\hat{\bar{2}},\widetilde{(ij)},k)\ M_{4}^{0}(\hat{1},\hat{\bar{\bar{2}}},\widetilde{((ij)k)},l)\nonumber\\
 &-&f_{3}^{0}(i,j,k)\ F_{3}^{0}(\hat{2},\widetilde{(ij)},\widetilde{(jk)})\ M_{4}^{0}(\hat{1},\hat{\bar{2}},\widetilde{((ij)(jk))},l)\nonumber\\
 &-&f_{3}^{0}(j,k,\hat{2})\ F_{3}^{0}(\hat{\bar{2}},i,\widetilde{(jk}))\ M_{4}^{0}(\hat{1},\hat{\bar{\bar{2}}},\widetilde{(i(jk))},l)
\end{eqnarray}\vspace{-0.5cm}
\newline where $f_{3}^{0},F_{3}^{0},F_{4}^{0}$ are the various three and four gluon antenna functions, and their arguments reflect the parton momenta generated via the appropriate phase space maps~\cite{Kosowermap,initialnlo}.  This process was considered in \cite{gluons} and such a block of subtraction terms can be found in the appropriate double-real subtraction term formula. The generality of this approach can be seen by considering a different physical process, for example a quark-antiquark pair scattering into four gluons where we consider the colour ordered six parton matrix elements $M_{6}^{0}(\hat{1}_{q},i,j,k,l,\hat{2}_{\bar{q}})$.  In this case the block of diagrams in Fig. \ref{w4block} translates to a subtraction term given by,
\begin{eqnarray}
D_{4}^{0}(\hat{1}_{q},i,j,k)\ M_{4}^{0}(\hat{\bar{1}}_{q},\widetilde{(ijk)},l,\hat{2}_{\bar{q}})&-& d_{3}^{0}(\hat{1}_{q},i,j)\ D_{3}^{0}(\hat{\bar{1}}_{q},\widetilde{(ij)},k)\ M_{4}^{0}(\hat{\bar{\bar{1}}}_{q},\widetilde{((ij)k)},l,\hat{2}_{\bar{q}})\nonumber\\
 &-&f_{3}^{0}(i,j,k)\ D_{3}^{0}(\hat{1}_{q},\widetilde{(ij)},\widetilde{(jk)})\ M_{4}^{0}(\hat{\bar{1}}_{q},\widetilde{((ij)(jk))},l,\hat{2}_{\bar{q}})\nonumber\\
 &-&d_{3}^{0}(\hat{1}_{q},k,j)\ D_{3}^{0}(\hat{\bar{1}}_{q},i,\widetilde{(kj}))\ M_{4}^{0}(\hat{\bar{\bar{1}}}_{q},\widetilde{(i(kj))},l,\hat{2}_{\bar{q}})
\end{eqnarray}\vspace{-0.4cm}
\newline where $d_{3}^{0},D_{3}^{0},D_{4}^{0}$ are quark-gluon antenna functions defined in \cite{NNLOantenna}.  By deriving blocks of subtraction terms in this manner and carefully keeping track of any over-subtraction, the full subtraction term can be formulated systematically.  The details of the subtraction term for each partonic channel vary according to the matrix elements and antenna functions involved but can be derived within this basic framework.  In order to keep track of the parton species involved in a process it is often convenient to draw the polygon edges to reflect the parton it represents as in Fig. \ref{newedges}; in this way the antenna species and specific singular limits can be read off the diagrams directly.
\vspace{-0.6cm}
\section{Conclusions}
In this talk, I have outlined the general procedure by which single and double unresolved IR singularities can be isolated using the antenna subtraction method.  The method relies on two factorization theorems: antenna phase space factorization and universal matrix element factorization.  These properties are exploited to construct a subtraction term which mimics the double-real radiation cross section in the singular limits and thus absorbs all divergences without introducing additional spurious singular behaviour.  A novel pictorial method is introduced for constructing the double-real radiation subtraction term within the antenna subtraction formalism.  The method allows general blocks of subtraction terms to be generated using only kinematic information; these can then be dressed with process specific information to determine the final form of the subtraction formula.

\section{Acknowledgements}
I gratefully acknowledge useful discussions with Thomas Gehrmann, Aude Gehrmann-De Ridder, Nigel Glover and Joao Pires.  This work is supported by the Science and Technologies Facilities Council.

\bibliography{radcor}

\end{document}